\documentclass{article}
\usepackage[pdftex]{color}
\usepackage{a4wide,amssymb,amsmath,paralist,epsfig,epstopdf}
\usepackage[utf8]{inputenc}
\DeclareGraphicsExtensions{.pdftex,.pdf}

\newtheorem{lemma}{Lemma}
\newtheorem{theorem}{Theorem}

\newenvironment{myproof}{{\it Proof:\/}}{\hspace*{1em}\hfill$\Box$\vskip 0.1in}

\begin{document}

\title{Reconstructing 3-colored grids from\\
 horizontal 
  and vertical projections is NP-hard}

\author{Christoph D\"urr%
\thanks{CNRS, LIX (UMR 7161),
  Ecole Polytechnique,
  91128 Palaiseau, France.}
\and
Flavio Gui\~nez%
\thanks{DIM, Universidad de Chile, Casilla 170-3, Correo 3, Santiago, Chile.}
\and
Mart\'{\i}n Matamala%
\thanks{DIM and CMM (UMI 2807, CNRS), Universidad de Chile, 
  Casilla 170-3, Correo 3, Santiago, Chile.}
}

\maketitle


\begin{abstract}
We consider the problem of coloring a grid using $k$ colors with the
restriction that in each row and each column has an specific number of
cells of each color.  In an already classical result, Ryser obtained a
necessary and sufficient condition for the existence of such a
coloring when two colors are considered. This characterization yields
a linear time algorithm for constructing such a coloring when it
exists.  Gardner \emph{et al.} showed that for $k\geq 7$ the problem
is NP-hard. Afterward Chrobak and D\"urr improved this result, by
proving that it remains NP-hard for $k\geq 4$. We solve the gap by
showing that for $3$ colors the problem is already NP-hard.  Besides
we also give some results on tiling tomography problems.
\end{abstract}

\section{Introduction}

Tomography consists of reconstructing spatial objects from lower
dimensional projections, and has medical applications as well as
non-destructive quality control.  In the discrete variant, the objects
to be reconstructed are discrete, as for example atoms in a crystaline structure, see~\cite{Alpers.Rodek.ea:Advances-in-Discrete}.

One of the first studied problem in discrete tomography involves the
coloring of a grid, with a fixed number of colors with the requirement
that each row and each column has a specific total number of entries
of each color.
 
More formally we are given a set of colors $\cal C$, and an $m\times
n$ matrix $M$, whose items are elements of $\cal C$.  The projection
of $M$ is a sequence of vectors $r^c\in{\mathbb N}^m,s^c\in{\mathbb
  N}^n$, for $c\in\cal C$, where
\begin{align*}
  r^c_i &= |\{ j: M_{ij}=c \}|,
&
  s^c_j &= |\{ i: M_{ij}=c \}|.
\end{align*}
In the reconstruction problem, we are given only a sequence of vectors
satisfying for $1\leq i\leq m$, $1\leq j\leq n$,  $c\in\cal C$,
\begin{align}   \label{eq:sum=ok}
  \sum_c r^c_i &=n, &
  \sum_c s^c_j &=m, &
  \sum_i r^c_i &= \sum_j s^c_j,
\end{align}
 and the goal is to compute \emph{a} matrix $M$ that has the given
 projections.  If there are $k=|{\cal C}|$ colors, we call it the
 \textsc{$k$-color Tomography Problem}.  

It was known since long time, that for 2 colors, the problem can be
solved in polynomial time~\cite{ryser60:_matric_zeros_ones}.  Ten
years ago it was shown that the problem is NP-hard for 7
colors~\cite{GaGrPr99}.  By NP-hardness, we mean that the decision
variant --- deciding whether a given instance is \emph{feasible},
i.e.\ admits a solution --- is NP-hard.  Shortly after this proof was
improved to show NP-hardness for 4 colors, leaving open the case when
$|{\cal C}|=3$~\cite{ChrobakDurr01}.  This paper closes the gap, by
showing that for 3 colors already the problem is NP-hard.

Just to fix the notation, for $|{\cal C}|=2$ we denote the colors as
\emph{black} and \emph{white}, and use symbols $B,W$.  For $|{\cal
C}|=3$ we denote the colors as \emph{red}, \emph{green} and
\emph{yellow} and use symbols $R,G,Y$.  Notice that we can think white
and yellow as ground colors in the $2$ and $3-$color problem,
respectively. Thus when we denote the instance of the tomography
problem, we sometimes omit the white or yellow projections as they are
redundant.  In addition for a 2-color instance $(r^B,s^B)$ we omit the
superscript when the context permits it.

First we recall some well known facts about the \textsc{2-color tomography
problem}.
\begin{lemma}[\cite{ryser60:_matric_zeros_ones}]   \label{lem:separation}
  Let $(r,s)$ be a feasible instance of the 2-color tomography
  problem.  Let $I$ be some set of rows, and $J$ be some set of
  columns.  If
\begin{equation} \label{eq:IJ}
   \sum_{i\in I} r_i - \sum_{j\not\in J} s_j = |I\times J|,
\end{equation}
  then every solution to the instance will be all black in $I\times J$
  and all white in $\overline I \times \overline J$.
\end{lemma}
\begin{myproof}
  The sets $I$,$J$ divide the grid into four parts, $I\times J$,
  $I\times \overline J$, $\overline I \times J$ and $\overline I
  \times \overline J$.  The value $\sum_{i\in I} r_i$ equals the
  number of black cells in the first two parts, and $\sum_{j\not\in J}
  s_j$ the number of black cells in the second and last part.  So the
  difference is the number of black cells in $I\times J$ minus the
  number of black cells in $\overline I \times \overline J$.  So when
  (\ref{eq:IJ}) holds, the first part must be all black and the last
  part all white.
\end{myproof}

\begin{figure}[htb]
  \centerline{\includegraphics[width=2cm]{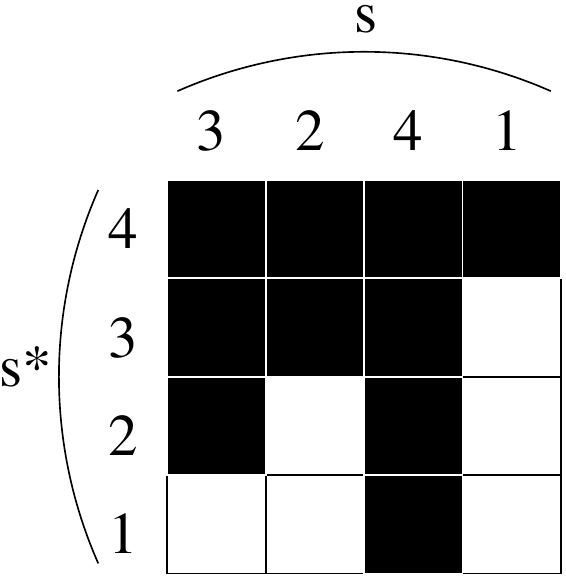}}
  \caption{Example of a vector $s$ and its conjugate $s^*$.}
  \label{fig:conjugate}
\end{figure}

Before stating the next lemma, we need to introduce some notation
about vectors.  The \emph{conjugate} of a vector
$s\in\{0,1,\ldots,m\}^n$ is defined as the vector
$s^*\in\{0,1,\ldots,n\}^m$ where $s^*_i = |\{ j : s_j \geq i \}|$.
There is a very simple graphical interpretation of this.  Let be an
$m\times n$ matrix $M$, such in column $j$, the first $s_j$ cells are
colored black and the others are colored white.  Then the conjugate of
$s$ is just the row projection of $M$, see figure~\ref{fig:conjugate}.

Note that $s^*$ is always a non-increasing vector.  If in addition $s$
is non-increasing we have that $(s^*)^*=s$ since in this case
$s^*_i=\max\{j: s_j\geq i\}$ and $s^*_i\geq j$ if and only if $s_j\geq
i$.  

For every $s,t\in \mathbb N^n$ we say that $s$ dominates $t$, denoted
$s\succeq t$, if $\sum_{j=1}^\ell s_j\geq \sum_{j=1}^\ell t_j$ for
every $1\leq \ell\leq n$.  For any $0\le k\le n$ we define the set
${\cal X}_{n,k} := \{ x\in\{0,1\}^n : \sum x_i = k \}$.  Clearly
$\succeq$ defines a partial order on ${\cal X}_{n,k}$, and we show now
that it has a \emph{small depth}.
\begin{lemma}[\cite{ChrobakDurr01}]      \label{lem:depth-dom}
  Let $n,k$ be two integers with $0\leq k\leq n$.  Suppose we have a
  strictly increasing sequence
\[
   b^0 \prec b^1 \prec \ldots \prec b^q,
\]
  of vectors from ${\cal X}_{n,k}$. Then $q \leq k(n-k)$.
\end{lemma}
\begin{myproof}
  For each vector $\alpha\in{\cal X}_{n,k}$ we associate the number
  $\varphi(\alpha)$ defined by $\varphi(\alpha) = \sum_{\ell=1}^n
  \sum_{i=1}^\ell \alpha_i$. 

  If $\alpha \prec \beta$ then $\sum_{j=1}^\ell \alpha_j\leq
  \sum_{j=1}^\ell \beta_j$ for every $1\leq \ell\leq n$ and the
  inequality is strict for at least one $\ell$.  We conclude that
  $\alpha \prec \beta$ implies $\varphi(\alpha) < \varphi(\beta)$.

  Therefore the vectors with extreme values for $\varphi$ are
  $\alpha=(0,\ldots,0,\underbrace{1,\ldots,1}_k)$ and
  $\beta=(\underbrace{1,\ldots,1}_k,0,\ldots,0)$.  Since
  $\varphi(\alpha)=k(k-1)/2$ and $\varphi(\beta)=k(k-1)/2+k(n-k)$,
  this concludes the proof.
\end{myproof}

A well-known characterization of the feasible instances of the
\textsc{2-color tomography problem} can be expressed using
dominance.

\begin{lemma}[\cite{ryser60:_matric_zeros_ones}]  \label{lem:dominance}
  Let $(r,s)$ be an instance of the 2-color tomography problem, such
  that $r$ is non-increasing.  Then $(r,s)$ is feasible if and only if
  $r \preceq s^*$.  Moreover if $r= s^*$, then there is a single
  solution, namely the realization having the first $s_j$ cells of
  column $j$ colored black, and the others white.
\end{lemma}
There is a very simple graphical interpretation of this.  Again let
$M$ be a matrix where in column $j$ the first $s_j$ cells are colored
black and the remaining cells white.  Then the row projection of $M$
is $s^*$, and if $s^*=r$ we are done.  Now if $s^* \neq r$, then some
of the black cells in $M$ have to be exchanged with some white cells
in the same column but a lower row.  These operations transform the
matrix in such a way, that the new row projection is dominated by
$s^*$.  So if $s^*$ does not dominate $r$, then there is no solution
to the instance.

\section{The gadget}
  The gadget depends on some integers $n,k,u,v$ with $1\leq k,u,v \leq
  n$ and $u\neq v$ as well as on two vectors $\alpha,\beta\in{\cal
    X}_{n,k}$.  It is defined as the instance of $n$ rows, and $2n+2$
  columns with the following projections for $1\leq i,j \leq n$
  \begin{align*}
    r^R_{i} &= \left\{ 
    \begin{array}{ll}
      i +1  &\mbox{if } i \in \{u,v\} \\
      i &\mbox{otherwise}
    \end{array}
    \right.
    &
    r^G_{i} &= \left\{ 
    \begin{array}{ll}
     i  &\mbox{if } i \in \{u,v\} \\
     i+1 &\mbox{otherwise}
    \end{array}
    \right.
    \\[2em]
    s^R_j &= n-j+\alpha_j
    &
    s^G_j &= 0
    \\
    s^R_{n+1} &= 1 
    &
    s^G_{n+1} &= n-1
    \\
    s^R_{n+2} &= n-k+1
    &
    s^G_{n+2} &= k-1
    \\
    s^R_{n+2+j} &= 0
    &
    s^G_{n+2+j} &= n-j+1-\beta_j.
  \end{align*}
\begin{lemma} \label{lem:gadget}
  If the instance above is feasible then $\alpha \preceq \beta$.
  Moreover, if $\alpha=\beta$ then the instance is feasible if and
  only if $\alpha_u+\alpha_v \geq 1$.
\end{lemma}
\begin{myproof}
  Assume the instance is feasible, we will show that this implies
  $\alpha \preceq \beta$.  Consider the yellow projection vectors
  $r^Y=2n+2-r^R-r^G$ and $s^Y=n-s^R-s^G$. We have that $r^Y_i =
  2(n-i)+1$ for $1\leq i \leq n$.  Note that $r^Y$ is a non-increasing
  vector.  Similarly, we obtain that $s^Y_j = j-\alpha_j$ and
  $s^Y_{n+2+j} = j-1+\beta_j$, for $1\leq j \leq n$, and
  $s^Y_{n+1}=s^Y_{n+2}=0$.  The conjugate of the column yellow
  projections is a vector $(s^Y)^*$ with
\[
     (s^Y)^*_i = 2(n-i)+1-\alpha_i+\beta_i.
\]
  Then clearly $r^Y \preceq (s^Y)^*$ if and only if $\alpha \preceq
  \beta$.  By assumption the 3-color instance
  $(r^R,r^G,r^Y,s^R,s^G,s^Y)$ is feasible, therefore the 2-color
  instance $(r^Y,s^Y)$ is feasible as well --- where yellow is renamed
  as black --- which by Lemma~\ref{lem:dominance} implies $r^Y \preceq
  (s^Y)^*$ and therefore also $\alpha \preceq \beta$.  This shows the
  first part of the lemma.

  Now assume that the instance has a solution, and $\alpha=\beta$.
  The $n \times (2n+2)$ grid is divided into 3 parts (see
  figure~\ref{fig:gutterGadget+full}): into an $n\times n$ block
  (called \emph{RY-block}), a $n\times 2$ rectangle (called
  \emph{2-column translator}) and another $n\times n$ block (called
  \emph{GY-block}).  Again every block is sub-divided into an upper
  triangle, a diagonal and a lower triangle.

\begin{figure}[htb]
  \centerline{\includegraphics[width=12cm]{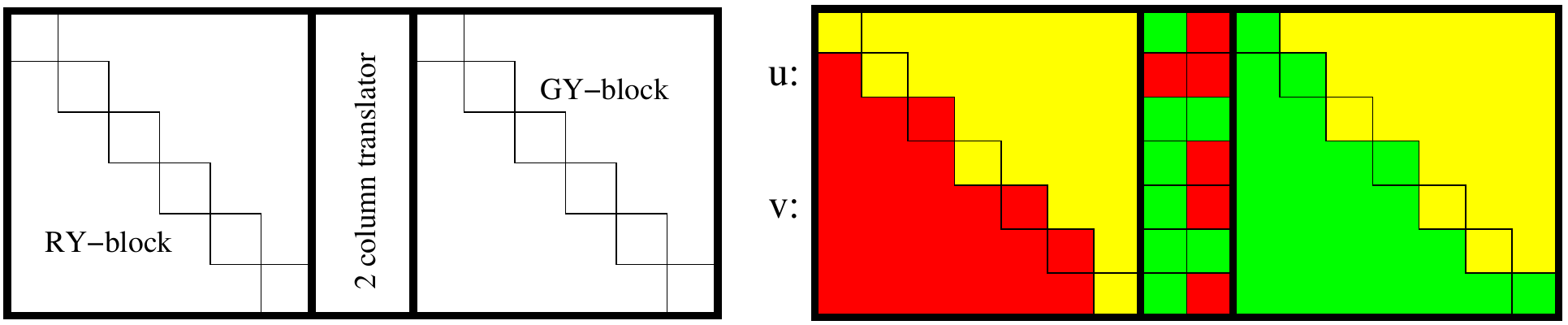}}
  \caption{The structure of the gadget (left) and a realization
  (right) for $n=7$, $k=3$, $u=2$, $v=5$ and $\alpha=\beta=(0,0,1,0,1,1,0)$.}
  \label{fig:gutterGadget+full}
\end{figure}

  Since $\alpha=\beta$, we have $r^Y = (s^Y)^*$. So by
  Lemma~\ref{lem:dominance} any solution must color in yellow the
  $s^Y_j$ first cells in every column $j$, and no other cell.  In
  particular it means that the lower triangle of the RY-block must be
  red, the lower triangle of the GY-block must be green, and both
  upper triangles have to be yellow.

  Also on the first diagonal, the cell $(i,i)$ has to be red if
  $\alpha_i=1$ and yellow otherwise.  On the second diagonal, the cell
  $(n+2+i,i)$ must be yellow if $\alpha_i=1$ and green otherwise.

  What can we say about the colors of the translator?  If
  $\alpha_u=\alpha_v=0$, then the cells
  $(n+1,u),(n+2,u),(n+1,v),(n+2,v)$ have to be all red to satisfy the
  row projections.  This contradicts the column projection
  $s^R_{n+1}=1$, and hence the instance is not feasible.

  Conversely, assume $\alpha_u+\alpha_v\geq 1$.  We will color the
  cells of the translator in a manner that respects the required
  projections.  If $i\not\in\{u,v\}$ and $\alpha_i=1$ --- that is
  $(i,i)$ is red --- we color the cells $(n+1,i),(n+2,i)$ in green. If
  $i\not\in\{u,v\}$ and $\alpha_i=0$, we color the cell $(n+1,i)$ in
  green and $(n+2,i)$ in red.

  Without loss of generality assume that $\alpha_u=1$. Hence $(u,u)$
  is red and we color $(n+1,u)$ in green and $(n+2,u)$ in red. We
  color $(n+1,v)$ in red. In addition we color $(n+2,v)$ in red if
  $\alpha_v=0$ and in green otherwise.  It can be verified that the
  coloring defined above is a solution to the instance, which
  concludes the proof of the lemma.
\end{myproof}

\section{The reduction}

In this section we will construct a reduction from \textsc{Vertex
Cover} to \textsc{3-color tomography}.  We basically use the same
approach than in~\cite{ChrobakDurr01}, but with a different gadget.

\textsc{Vertex Cover}
is a well known intractable problem, indeed one of the first 21
problems shown to be NP-complete by Karp~\cite{GarJoh79}.

\vspace{.2cm}
\textsc{Vertex Cover Problem}
\begin{itemize}
\item[] \textbf{Input:} a graph $G=(V,E)$ and an integer $k$.
\item[] \textbf{Output:} a set $S\subseteq V$ of size $|S|=k$ such
that for every $(u,v) \in E$, $u\in S$ or $v\in S$.
\end{itemize}
\vspace{.2cm}

Given an instance $(G,k)$ of \textsc{Vertex Cover}, we construct an
instance $(r^R,r^B,s^R,s^B)$ of the \textsc{3-color tomography
problem} which is feasible if and only if the former instance has a
solution.  Without loss of generality we assume that $k\leq n-2$.

Let be $n=|V|, m=|E|$, and $N=k(n-k)(m-1)+1$.  We denote the $m$ edges
as $E=\{e_0,e_1,\ldots,e_{m-1}\}$, and the $n$ vertices as
$V=\{1,2,\ldots,n\}$.  We define an instance with $N(n+1)+1$ rows
and $N(n+2)+n$ columns.


For row $p=1,\ldots,N(n+1)+1$, let 
\begin{align*}
x &= \lfloor (p-1)/(n+1) \rfloor\\
i &= (p-1) \bmod (n+1).
\end{align*}
We think the set of rows as divided into $N$ blocks of $n+1$ rows
each, and a last block with a single row.  We have $x$ as the block
index and $i$ the row index relative to the block, with $0\leq i\leq
n$.  Let $t= x \bmod m$ and consider the edge $e_t=(u,v)$. We define
the projections
\begin{align*}
  r^R_p &= x(n+2) + \left\{ \begin{array}{ll}
                n-k  &\mbox{if } x<N \mbox{ and } i=0 \\
	   	0    &\mbox{if } x=N \mbox{ and } i=0 \\
	       	i+1  &\mbox{if } i \in \{u,v\} \\
		i    &\mbox{if } i \in \{1,\ldots, n\}\setminus \{u,v\} \\
  \end{array}
  \right.
\\
  r^G_p &= (N-x-1)(n+2) + \left\{ \begin{array}{ll}
                n+2      &\mbox{if } x=0 \mbox{ and } i=0 \\
	        n+2+k    &\mbox{if } x>0 \mbox{ and } i=0 \\
		i        &\mbox{if } i \in \{u,v\} \\
		i+1      &\mbox{if } i \in \{1,\ldots, n\}\setminus \{u,v\}.\\
	   	   \end{array}
  \right.
\end{align*}


In the same manner, for column $q=1,\ldots,N(n+2)+n$, let 
\begin{align*}
y &= \lfloor (q-1) / (n+2) \rfloor\\
j &= ((q-1) \bmod (n+2))+1.
\end{align*}
The reason for defining $j$ this way, is that if cell $(p,q)$ is part of an RY-block or an GY-block, then $(i,j)$ will be the relative position inside the block with ranges $1\leq i,j\leq n$.
Similarly as for the rows, we think the set of columns as divided into
$N$ blocks with $n+2$ columns each and a last block with only $n$
columns.  Again, we have $y$ as the block index, and $j$ as the column
index relative to a block with $1\leq j \leq n+2$.
  For block $0 \leq y\leq N-1$ we
define the red column projections as
\begin{align*}
  s^R_q &= (N-y-1)(n+1)+1+
  \left\{ \begin{array}{ll}
    n-j+1   &\mbox{if } j\in\{1,\ldots, n\} \\
    1       &\mbox{if } j=n+1 \\
    n-k+1   &\mbox{if } j=n+2, \\
    		   \end{array}
  \right.
\end{align*}
For $y=N$ we set $s^R_q=0$, for each $j=1,\ldots,n$.  Similarly, for
$y=0$ the green column projections are
\begin{align*}
  s^G_q &= 
  \left\{ \begin{array}{ll}
    0       &\mbox{if } j\in\{1,\ldots, n\} \\
    n       &\mbox{if } j=n+1 \\
    k       &\mbox{if } j=n+2.\\
    		   \end{array}
  \right.
\end{align*}
and for $1\leq y\leq N$ they are defined as
\begin{align*}
  s^G_q &= 
  (y-1)(n+1)+1+
  \left\{ \begin{array}{ll}
    j   &\mbox{if } j\in\{1,\ldots, n\} \\
    n -1    &\mbox{if } j=n+1 \\
    k-1     &\mbox{if } j=n+2.\\
    		   \end{array}
  \right.
\end{align*}

\begin{figure}[htb]
  \centerline{\includegraphics[width=10cm]{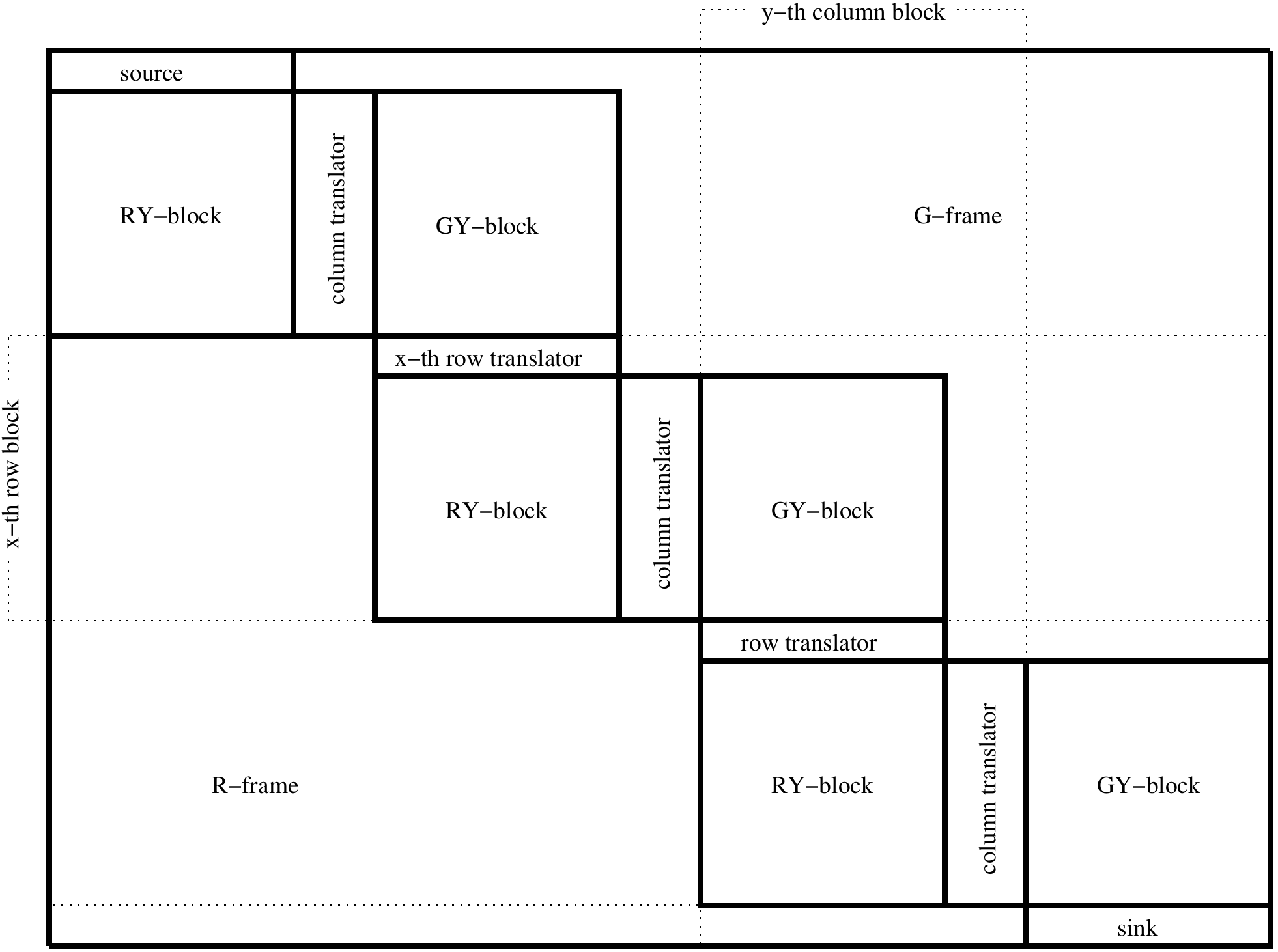}}
  \caption{The general structure of our reduction}
  \label{fig:gutter}
\end{figure}

Clearly this a polynomial time reduction.  It remains to show the
following theorem.
\begin{theorem}
  The 3-color tomography instance is feasible if and only if the
  vertex cover instance is feasible.
\end{theorem}
\begin{myproof}
  For one direction of the statement, assume that the vertex cover
  instance is feasible, and let $b\in{\cal X}_{n,k}$ be the
  characteristic vector of a vertex cover of size $k$, i.e.\
  $b_i=1$ if and only if $i$ belongs to the vertex cover.
  
  We construct now a solution to the tomography instance. Consider the
  partitioning of the grid, as in figure~\ref{fig:gutter}.  For
  convenience we refer to the source also as the $0$-th row translator
  and to the sink as the $(N+1)$-th row translator.  The $j$-th cell of
  the $x$-th row translator is defined as $(x(n+1)+1,x(n+2)+j)$.  We
  color the R-frame in red and the G-frame in green.

  Let be any $1\leq j \leq n$.  We color the $j$-th cell of the source
  in yellow if $b_j=1$ and in red otherwise.  For $x=1,\ldots,N-1$ we
  color the $j$-cell of the $x$-th row translator in green if $b_j=1$
  and in red otherwise. In the sink we color the $j$-th cell in green
  if $b_j=1$ and in yellow otherwise.

  Now for block $x=0,\ldots,N-1$, consider the instance to the gadget
  defined by $\alpha=\beta=b$, and $u,v$ such that $(u,v)=e_{x \bmod
    m}$.  By Lemma~\ref{lem:gadget} it is feasible, since $b$ is a
  vertex cover and hence $b_u+b_v\geq 1$.  Then we color the
  $(n+1)\times(2n+2)$ cells starting at $((x-1)(n+2)+1,(x-1)(n+1)+1)$
  exactly as in the solution to the gadget.  It is straightforward to
  check that this grid satisfies the required projections, and
  therefore the tomography instance is feasible.

  For the converse, assume that the tomography instance has a
  solution.  For every $x=1,\ldots,N$ we apply
  Lemma~\ref{lem:separation} for the red color and intervals
  $I=[x(n+1)+1, N(n+1)+1]$ and $J=[1,x(n+2)]$. We deduce that in the
  solution the R-frame must be all red, and all GY-blocks (and also
  the G-frame) must be free of any red.  Similarly, we show that the
  G-frame must be all green, and all RY-blocks must be free of any
  green.

  This implies that in the source, $k$ cells are yellow, and $n-k$ are
  red, in the row translators $k$ cells are green and $n-k$ red, and
  in the sink $k$ cells are green and $n-k$ yellow.  We define the
  vectors $b^0,b^1,\ldots, b^N \in {\cal X}_{n,k}$, such that for all
  $1\leq j\leq n$ we have
  \begin{itemize}
    \item $b^0_j=1$ iff the $j$-th cell in the source is yellow,
    \item $b^x_j=1$ iff the $j$-th cell in the $x$-th row translator is green,
      for all $1\leq x\leq N$.
  \end{itemize}

  For $x=0,\ldots,N$, consider the part $P$ of the solution that is
  the intersection of rows $[x(n+1)+2,x(n+1)+n+1]$ and columns
  $[x(n+2)+1, x(n+2)+2n+2]$.  We number the rows of $P$ from $1$ to
  $n$ and the columns from $1$ to $2n+2$.  Let $(u,v)=e_{x\bmod m}$.
  By subtracting from the row projections the number of red and green
  cells in the frames, we deduce that row $1\leq i \leq n$ in $P$
  contains $i+1$ red cells and $i$ green cells if $i\in\{u,v\}$ and
  $i$ red cells and $i+1$ green cells if $i\not\in\{u,v\}$.

  We proceed similarly for the columns $n+1$ and $n+2$.  By
  subtracting from the column projections the quantities that are in
  the frames, we deduce that column $n+1$ of $P$ contains one red
  cell, and $n-1$ green cells, and column $n+2$ contains $n-k+1$ red
  cells and $k-1$ green cells.

  Column $x(n+2)+j$ for $1\leq j \leq n$ contains $n-j+1$ red cells
  that are not in the R-frame.  Since GY-blocks are free of red, these
  cells must either be in the $x$-th row translator or in column $j$
  of $P$.  Note that the $j$-cell of the $x$-th row translator is red
  iff $b^x_j =0$.  Therefore column $j$ of $P$ contains $n-j+b^x_j$
  red cells and no green cell.  Similarly column $n+2+j$ of $P$
  contains $n-j+1-b^{x+1}_j$ green cells and no red cell.

  This implies that $P$ is the solution to the gadget defined by
  $u,v,\alpha,\beta$ with $\alpha=b^x$ and $\beta=b^{x+1}$.  Then by Lemma~\ref{lem:gadget} we obtain that $b^x \preceq b^{x+1}$ and in general
\[
    b^0 \preceq b^1 \preceq \ldots \preceq b^N.
\]
  By the choice of $N$ and Lemma~\ref{lem:depth-dom} there exists an $\ell$
  such that
\[
    b^{\ell} = b^{\ell+1} = \ldots = b^{\ell+m}.
\]
  By Lemma~\ref{lem:gadget}, we have $b^{\ell}_u+b^{\ell}_v \geq 1$
  for all $(u,v)\in\{ e_\ell, e_{\ell+1 \bmod m},
  \ldots, e_{\ell+m-1 \bmod m} \} = E$.  Therefore $b^{\ell}$ encodes
  a vertex cover of size $k$, and this completes the proof.
\end{myproof}


\section{Related problems}

\subsection{Edge-colored graphs with prescribed degrees}

We can reduce the \textsc{$3$-color tomography problem} to a similar
graph problem.  

\paragraph{Finding edge-colored graphs with prescribed degrees}
Let be a set of two colors $\{R,G\}$, and a vertex set $V$.  We are
given prescribed degrees $d^R, d^G : V \rightarrow \mathbb N$ and have
to find two disjoint edge sets $E^R, E^G \subseteq V^2$ such that the
graph $G(V,E^R \cup E^G)$ has the required degrees, i.e. for all $v\in
V$
\begin{align*}
    d^R(v) = | \{ u: (u,v)\in E^R \} | 
&&
    d^G(v) = | \{ u: (u,v)\in E^G \} |.
\end{align*}

Note that in contrast, finding an uncolored graph with given degree
sequences can be solved in polynomial time, see for example~\cite{Kuba.Herman:Discrete-tomography}.
\begin{lemma}
  The problem of finding an edge-colored graph with prescribed degrees
  is NP-hard.
\end{lemma}
\begin{myproof}
  We reduce from the $3$-color tomography problem.  Let $(r^R, r^G,
  s^R,s^G)$ be an $m\times n$-instance of the 3-color tomography
  problem.  We set $k=n+m$, $V=\{1,\ldots,k\}$, and the following
  degrees, for $1\le i\le n$ and $1\le j\le m$
\begin{align*}
      d^R(i)&=r^R_i + n-1 &
      d^G(i)&=r^G_i\\
      d^R(n+j)&=s^R_j  &
      d^G(n+j)&=s^G_j+m-1.\\
\end{align*}
  Now we show that the instance $(r^R, r^G, s^R,s^G)$ is feasible if
  and only if the instance $(d^R,d^G)$ is feasible.  For one
  direction, assume that there is a solution $M$ to the 3-color
  tomography instance.  We construct a solution $E^R,E^G$ to the graph
  problem as follows.  For any $1\le i\le n$ and $1\le j\le m$, if
  $M_{ij}=R$, then $(i,n+j)\in E^R$, if $M_{ij}=G$, then $(i,n+j)\in
  E^G$.  Also for any $1\le i<i' \le n$, we have $(i,i')\in E^R$ and
  for any $1\le j<j'\le m$, we have $(n+j,n+j')\in E^G$.  Now clearly
  $E^R,E^G$ satisfy the required degrees.

  For the converse, we define the quantity $\Phi=\sum_{i=1}^n
  d^R(i)-\sum_{j=1}^m d^R(n+j)$.  By assumption~(\ref{eq:sum=ok}) this
  value is $n(n-1)$.  Since this value equals also
\[
     | E^R \cap \{1,\ldots,n\}^2| - |E^R \cap \{n+1,\ldots,n+m\}|,
\]
  there is a red edge between every pair of vertices $(i,i')$ with
  $1\leq i<i'<n$, and no edge between every pair of vertices
  $(n+j,n+j')$ with $1\le j<j'\le m$.  Similarly we can show that
  there is a green edge between every pair of vertices $(n+j,n+j')$
  with $1\le j<j'\le m$.

  Now let $M$ be the $m\times n$ grid, with cell $(i,j)$ colored in
  red if $(i,n+j)\in E^R$, and in green if $(i,n+j)\in E^G$.  By the
  degree requirements, $M$ is a solution to the 3-color tomography
  instance.
\end{myproof}

\subsection{Tiling Tomography}

Tiling tomography was introduced in  \cite{CCDW01}, and it consists of
constructing a tiling that satisfies some given row and column
projections for each type of tiles we admit.

Formally a tile is a finite set $T$ of cells of the grid $\mathbb N
\times \mathbb N$, that are 4-connected, in the sense that the graph
$G(T,E)$ is connected for $E=\{((i,j),(i',j')) : |i-i'|+|j-j'|=1\}$.
By $T+(i',j')=\{(i+i',j+j'):(i,j)\in T\}$ we denote a copy of $T$ that
is shifted $i'$ units down and $j'$ units to the right.  We say that a
set of tiles is \emph{feasible} if they do not intersect.  In addition
we say that it tiles the $m\times n$ grid if its (disjoint) union
equals the set of all grid cells, and we refer it as a \emph{tiling}.

In the tiling tomography problem we are given a finite set of tiles
${\cal T}=\{T_1,\ldots,T_k\}$, and vectors $r^d\in{\mathbb
  N}^m,s^d\in{\mathbb N}^n$ for $1\leq d\leq k$.  The goal is to
compute a matrix $M\in\{0,1,\ldots,k\}$ such that the set
\[
      \bigcup_{1\leq d\leq k} \{T_d+(i,j) : 1\leq i\leq n,1\leq j\leq m
       \mbox{ such that }M_{ij}=d, \mbox{ for }1\leq d\leq k \}
\]
is a tiling of the $m\times n$ grid, with the projections
\begin{align*}
  r^d_i &= |\{ j: M_{ij}=d \}| 
&
  s^d_j &= |\{ i: M_{ij}=d \}|.
\end{align*}

By \emph{width} and \emph{height} of a tile $T$ we understand the size
of the smallest intervals $I,J$ such that $T \subseteq I\times J$.
This definition extends to set of tiles.  A tile $T$ is said to be
rectangular if for every $(i',j')$ such that $\{T,T+(i',j')\}$ is
feasible, we have that the width of $\{T,T+(i',j')\}$ is at least
twice the width of $T$ or the height of the set is at least twice the
height of $T$.

It was conjectured in~\cite{CCDW01}, that for $T_1$ being a single
cell and $T_2$ a non-rectangular tile, the $\{T_1,T_2\}$-tiling
tomography problem is NP-hard.  This question is still open and
intriguing.

\subsection{Rectangular tiles}

Consider two rectangular tiles, $T_1$ being a $p_1\times q_1$
rectangle and $T_2$ a $p_2\times q_2$ rectangle,
i.e. $T_c=\{0,\ldots,p_c-1\}\times \{0,\ldots,q_c-1\}$, for
$c\in\{1,2\}$.  What can be said about the complexity of the
$\{T_1,T_2\}$-tiling tomography problem?

If $\gcd(p_1,p_2)=d>1$, then clearly any solution
$\bar M\in\{0,1,2\}^{m\times n}$ to a $\{T_1,T_2\}$-tiling tomography
instance $(r^c,s^c)$, must satisfy that if $\bar M_{ij}\neq 0$, then $i
\bmod d =1$.  Therefore the $\{T_1,T_2\}$-tiling tomography problem
can be reduced to the $\{T'_1,T'_2\}$-tiling tomography problem, with
$T'_1$ being a $(p_1/d)\times q_1$ rectangle, and $T'_2$ a
$(p_2/d)\times q_2$ rectangle.  We omit the formal reduction, which is
straightforward.

From now on suppose that $\gcd(p_1,p_2)=\gcd(q_1,q_2)=1$.  We
distinguish the following cases, up to row-column symmetry.
\begin{itemize}
\item If $p_1=p_2=1$, that is the tiles are two horizontal bars of
  length $q_1$ and $q_2$, then the problem can be solved in polynomial
  time (Theorem \ref{alg}). We use an idea already present
  in~\cite{durr.goles.ea:03:tiling*}, where it is proven for $q_1=1$.
\item If $p_1=q_2=1$ and $p_2=q_1=2$, then the tiles are called
  dominoes, and again the problem can be solved in polynomial time,
  although with a more involved algorithm~\cite{thiant06:_const_et_recon_de_pavag_de_domin}.
\item If $p_1=q_2=1$, $p_2\geq2$ and $q_1\geq 3$ then the problem is
  open. The first author conjectures that the problem is NP-hard,
  while the other two conjecture that it could be solved in polynomial
  time with a similar approach as
  in~\cite{thiant06:_const_et_recon_de_pavag_de_domin}.
\item If $p_1,q_1\geq 2$, then the problem is NP-hard (Theorem \ref{tiling-np-hard}).  In
  \cite{CCDW01} the special case $p_1=q_1=2$, $p_2=q_2=1$ was
  related to the 3-color tomography problem, and it is therefore also
  NP-hard.  We generalize this reduction in section~\ref{sec:2rect}
\item If there is a third rectangular tile $T_3$, then for the tile
  set $\{T_1,T_2,T_3\}$ the problem is NP-hard, see section~\ref{sec:3rect}.
\end{itemize}


\subsection{An algorithm for vertical bars}

\begin{theorem} \label{alg}
The tiling tomography problem can be solved in polynomial time for two
rectangular tiles of dimensions $p_1\times 1$ and $p_2\times 1$.
\end{theorem}

\begin{myproof}
The algorithm is the simple greedy algorithm, as the one used in
\cite{durr.goles.ea:03:tiling*}.  It iteratively \emph{stacks} bars in
the matrix.

Formally the algorithm is defined like this.  We construct a matrix
$A\in\{0,1,2\}^{m\times n}$ with the required projections.  Initially
$A$ is all $0$.  We maintain a vector $v$ such that $v_j$ is the
minimal $i$ such that $A_{i,j}\neq 0$, and $v_j=m+1$ if column $j$ of
$A$ is all zero.  Initially $v_j=m+1$ for all $1\leq j\leq n$.  We
also maintaing vectors $\bar r^1,\bar r^2, \bar s^1, \bar s^2$, which
represent the remaining projections.  Initially they equal the given
projections of the instance.  The vectors $(v,\bar r^1,\bar r^2, \bar
s^1, \bar s^2)$ define a more general tiling problem, where in every
column $j$, only the first $v_j-1$ cells have to be tiled.

\begin{quote}
\textbf{The algorithm:} 
Let $i=\max v_j$.  If $i=1$ we are done, and return $A$, if all
vectors $\bar r^1,\bar r^2, \bar s^1, \bar s^2$ are zero, and return
``no solution'' otherwise.

If $i> 1$, let $i_1 = i -p_1$ and $i_2=i-p_2$.  If
$\bar r^1_{i_1}=\bar r^2_{i_2}=0$, abort and return ``no solution''.  Otherwise
let $c\in\{1,2\}$ such that $\bar r^c_{i_c} > 0$.  
Let $j$ be a column with $v_j+1=i$ that maximizes $\bar s^c_j$.
Then \emph{drop} the bar $p_c\times 1$ in column $j$, i.e.\ set $A_{i_c,j}=c$,
and decrease $\bar r^c_{i_c}$ and $\bar s^c_j$.  Repeat the whole step.
\end{quote}

Clearly, if this algorithm produces a matrix, then it defines a valid
tiling with the required projections.  We have to show that if the
instance has a solution, then the algorithm will actually find one.
For this purpose, let be some step of algorithm such that the
intermediate instance ${\cal I} := (v,\bar r^1,\bar r^2, \bar s^1,
\bar s^2)$ is feasible.  The initial step could be a candidate.  Let
$M$ be a solution to it.  Let $i=\max v_j$.  If $i=1$, then $\bar
r^1,\ldots, \bar s^2$ are all zero, since the instance is feasible.

Let $i_1 = i -p_1$ and $i_2=i-p_2$.  We have that either $M_{i_1,j}=1$
or $M_{i_2,j}=2$ for every column $j$ satisfying $v_j=i$, since $M$ is
a valid tiling.  Therefore some of $\bar r^1_{i_1}, \bar r^2_{i_2}$
must be non zero.  Let $c, j$ be the values the algorithm chooses.
Let ${\cal I'}$ be the instance obtained after the iteration of the
algorithm, that is $\bar r^c_{i_c},\bar s^c_{j}$ are decreased by $1$
and $v_j$ by $p_c$.

If $M_{i_c,j}=c$, then $M'$ which equals $M$ except for $M_{i_c,j}=0$
is a solution to ${\cal I'}$.

If $M_{i_c,j}\neq c$, then by the projections, there must be a another
column $k$ with $v_k=i$ and $M_{i_c,k}=c$.  We will now transform $M$
such that $M_{i_c,j}=c$.  Then we are in the case above and done.

By the choice of the algorithm we have $\bar s^c_k \leq \bar s^c_j$.
By this inequality, there exists $i_0$ such that the total number of
$c$'s below the row $i_0$ is the same in both column $j$ and column
$k$.  Take $i_0$ being the largest one satisfying that.  By the choice
of $i_0$ we have that $M_{i_0,j}=c$ and $M_{i_0,j}\neq c$.  Since $M$
is a valid tiling, then the restriction to cells below $i_0$ in column
$k$ is also a tiling and then $M_{i_0,j}\neq 0$. We conclude that
between $i_0$ and $i$ the number of 1's and 2's in column $j$ is the
same as in column $k$.  Then exchanging the parts of columns $j$ and
$k$ in $M$ between $i_0$ and $i$, does not change the projections of
$M$, and we obtain the required property $M_{i_c,j}=c$.

\begin{figure}
  \centerline{\includegraphics[width=8cm]{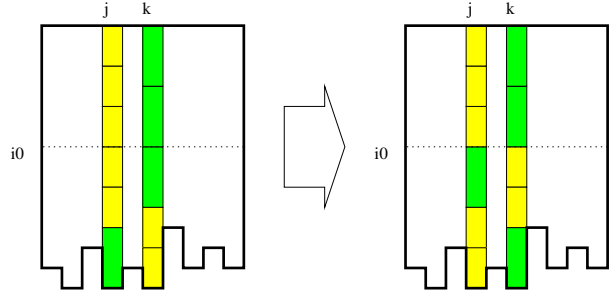}}
  \caption{Transforming a solution $M$ for $c=1$.}
  \label{fig:M}
\end{figure}

By the choice of the algorithm we have $s^c_k \leq s^c_j$.  Now we
claim that there is a row $i_0$ such between $i_0$ and $i$ the number
of 1's and 2's in column $j$ is the same as in column $k$.  Indeed,
consider the largest $i_0$ such that the total number of $c$'s below
the row $i_0$ is the same in both column $j$ and column $k$.  It must
be that $M_{i_0,j}=c$ and $M_{i_0,k}\neq 0$ since $M$ is a valid
tiling.  Then exchanging the parts of columns $j$ and $k$ in $M$
between $i_0$ and $i$, does not change the projections of $M$, and we
obtain the required property $M_{i_0,j}=c$.

\end{myproof}

\subsection{A general NP-hardness proof structure}   \label{sec:scheme}

In the next section we will reduce the 3-color tomography problem to
the tiling tomography problem for some fixed set of tiles $\cal T$.
The proof uses a particular structure that we explain now.

Let $(r^R,r^G,r^Y,s^R,s^G,s^Y)$ be an instance to the 3-color
tomography problem for an $m\times n$ grid.  In the reduction we will
choose constant size grid $\ell\times k$ --- that we call a
\emph{block} --- and three $\cal T$-tilings of it, that we denote
$\bar M^R,\bar M^G,\bar M^Y$.  There will be two requirements: Let
$\bar r^{c,d}, \bar s^{c,d}$ be the $T_d$-projections of the tiling
$\bar M^c$ for $c\in\{R,G,Y\}$ and $d\in\{1,2\}$.

\textbf{The first requirement} is that the vectors $\{\bar r^{R,1}, \bar
r^{G,1}, \bar r^{Y,1}\}$ are affine linear independent.  The same
requirement holds for the column projections $\{\bar s^{R,1}, \bar
s^{G,1}, \bar s^{Y,1}\}$.  This implies that every vector $r$ spanned
by $\bar r^{R,1}, \bar r^{G,1}, \bar r^{Y,1}$, has a unique
decomposition into $r=n_R\bar r^R+n_G\bar r^G+n_Y\bar r^Y$ for
$n_R+n_G+n_Y=n$.

The reduction, consists of an $m\ell\times nk$ grid, and the
projections $1\leq i\leq \ell$, $1\leq j\leq k$, $1\leq x\leq m$,
$1\leq y\leq n$, $d\in\{1,2\}$
\begin{align*}
    r^d_{x\ell-\ell+i} &= \sum_c r^c_{x} \cdot \bar r^{c,d}_{i}
\\ 
    s^d_{yk-k+j} &= \sum_c s^c_{y} \cdot \bar s^{c,d}_{j}
\end{align*}

The idea is that the $m\ell\times nk$ is partitioned into $mn$ blocks
of dimension $\ell\times k$. \textbf{The second requirement} is that
in every solution $\bar M$ to the tiling instance, all blocks of $\bar
M$, are either $\bar M^R,\bar M^G, \bar M^Y$ or blocks that have
equivalent projections.

\begin{lemma}                         \label{lem:tiling-structure}
  The instance to the $\cal T$-tiling problem has a solution if and
  only if the instance to the 3-color tomography problem has a
  solution.
\end{lemma}
\begin{myproof}
  Let $M\in\{R,G,Y\}^{m\times n}$ be a solution to the 3-color
  tomography problem.  We transform it into a matrix $\bar
  M\in\{0,1,2\}^{m\ell\times nk}$ by replacing each cell $(i,j)$ of
  $M$ by the $\ell\times k$ matrix $\bar M^c$ for $c=M_{ij}$.  By
  construction, this is a solution to the tiling problem.

  For the converse, suppose that there is a solution $\bar M$ to the
  tiling problem.  By the second requirement, every block of $\bar M$
  can be associated to one of the colors $\{R,G,Y\}$.  We construct a
  matrix $M\in\{R,G,Y\}^{m\times n}$ such that $M_{xy}=c$ if the block
  $(x,y)$ of $\bar M$ is $\bar M^c$, or something projection
  equivalent.  

  Fix some arbitrary $1\leq x\leq m$.  By the first requirement, the
  projections of the rows $x\ell-\ell+1,\ldots,x\ell$ have a unique
  decomposition into $n_R r^{R,1} + n_G r^{G,1} +n_Y r^{Y,1}$ with
  $n_R+n_G+n_Y=n$.  By the definitions of the projections $n_R =
  r^R_x, n_G=r^G_x, n_Y = r^Y_x$, and then row $x$ of $M$ has the required
  projections.  We proceed in the same manner for the columns and show
  that $M$ is a solution to the 3-color tomography instance.
\end{myproof}


\subsection{An NP-hardness proof for two rectangular tiles}
\label{sec:2rect}

\begin{theorem} \label{tiling-np-hard}
  The tiling tomography problem is NP-hard for two rectangular tiles
  of dimensions $p_1\times q_1$ and $p_2\times q_2$ with
  $\gcd(p_1,p_2)=\gcd(q_1,q_2)=1$ and $p_1,q_1 \geq 2$.
\end{theorem}
\begin{myproof}
  We apply Lemma~\ref{lem:tiling-structure} for $\ell=2p_1p_2$ and
  $k=2q_1q_2$.  The 3 tilings of the $\ell\times k$ grid are depicted
  in figure~\ref{fig:rectangles}, and defined formally as follows.
  The rows $I=\{1,\ldots,\ell\}$ and the columns $J=\{1,\ldots,k\}$
  are partitioned into sets $I_1,I_2,I_3,I_4$ and $J_1,J_2,J_3,J_4$
  defined as
\begin{align*}
     I_1 &= \{1,\ldots,p_2\} 
&    J_1 &= \{1,\ldots,q_2\} 
\\
     I_2 &= \{p_2+1,\ldots,p_1p_2\}
&    J_2 &= \{q_2+1,\ldots,q_1q_2\}
\\
     I_3 &= \{p_1p_2+1,\ldots,p_1p_2+p_2\}
&    J_3 &= \{q_1q_2+1,\ldots,q_1q_2+q_2\}
\\
     I_4 &= \{p_1p_2+p_2+1,\ldots,2p_1p_2\}
&    J_4 &= \{q_1q_2+q_2+1,\ldots,2q_1q_2\}.
\end{align*}
Then $\bar M^R$ is defined as the block tiling that covers $(I_1\cup
I_4)\times(J_3\cup J_4)$ with $T_2$ and the rest with $T_1$, $
\bar M^G$ is defined as the block tiling that covers $(I_3\cup
I_4)\times(J_1\cup J_4)$ with $T_2$ and the rest with $T_1$, while
$\bar M^Y$ is defined as a tiling using only $T_1$.  These tilings are
uniquely defined.  Clearly the row $T_1$-projections of the 3 tilings
are affine linear independent, so the first requirement of the
construction is satisfied.

  \begin{figure}
    \centerline{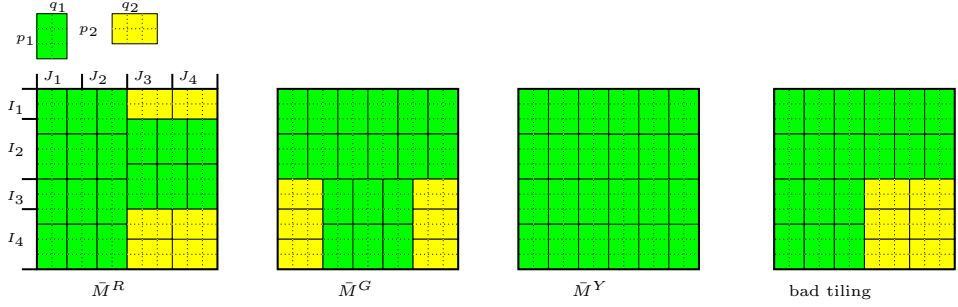}
    \caption{The 3 valid block tilings.}
    \label{fig:rectangles}
  \end{figure}

The second requirement follows from a sequence of observations.  Let
$\bar M$ be the solution to the tiling instance, obtained by reduction from
a 3-color instance $(r^R,r^G,r^Y, s^R,s^G,s^Y)$.

First note that in the tilings $\bar M^R,\bar M^G, \bar M^Y$, every
tile is completely contained in the $\ell\times k$ block.  Therefore
the tiling instance has zero projections for $T_1$ at rows $x$ with
$(x-1)\bmod \ell > \ell-p_1+2$.  A similar observation holds for tile
$T_2$ and for the column projections.  As a result in $\bar M$ every tile
is completely contained in some $\ell\times k$ block, and in other words every block
of $\bar M$ is $\{T_1,T_2\}$-tiled.  

What can we say about the possible
tilings?
Again note that in the tilings $\bar M^R,\bar M^G, \bar M^Y$, every
row in $I_2$ is completely covered by $T_1$-tiles.  Therefore by the
projections, this holds also for every block in $\bar M$.  The same
observation can be done about columns in $J_2$.

Note that if $ap_1 + bp_2 = 2p_1p_2$, then
$(a,b)\in\{(0,2p_1),(2p_2,0),(p_2,p_1)\}$.  This is simply because by
$\gcd(p_1,p_2)=1$, in any solution to $ap_1 = p_2(2p_1-b)$, $a$ must
be a multiple of $p_2$.  Together with the previous observation, this
implies that every column of a block is either covered completely by
$T_1$-tiles or covered half by $T_1$-tiles and half by $T_2$-tiles.  The
same observation holds for the rows.

The trickiest observation of this proof is that in every block of $\bar M$,
the region $I_1\times J_1$ is covered by $T_1$.  For a proof by
contradiction, suppose it is covered by $T_2$, in fact by a single tile $T_2$
since $|I_1\times J_1| = |T_2|$.  But since $I_2\times J$ is
covered with $T_2$, and by $\gcd(q_1,q_2)=1$, it must be that the cell
$(p_2+1,q_2+1)$ is covered by a tile $T_2+(p_2+1,j)$ for some column
$j\leq q_2$.  By the same argument, the cell $(p_2+1,q_2+1)$ is
\emph{also} covered by a tile $T_2+(q_2+1,i)$ for some row $i\leq
p_2$.  Therefore these two tiles overlap in $(p_2+1,q_2+1)$, which
contradicts that $M$ is a (valid) tiling.

Now fix a block of $\bar M$.  If row $1$ is partly covered by $T_2$,
then $T_2-$tiles must cover the half columns in $J$.  Hence in the row
$1$ they cover exactly the columns in $J_3\cup J_4$.  The same
argument shows that every column $j\in J_3\cup J_4$ is then half
covered by $T_2-$tiles.  Previous observation state that $I_2\times
\{j\}$ is covered by $T_1$.  But the length of $I_2$ is a not a
multiple of $p_1$.  Therefore $(p_1p_2+1,j)$ must then also be covered
by $T_1$ and hence $(I_2\cup I_3)\cup\{j\}$ is covered by
$T_1-$tiles. Therefore $(I_1\cup I_4)\times\{j\}$ is covered by $T_2$.
The choice of $j$ was arbitrary, and therefore the block-tiling is
exactly $\bar M^R$.

Similarly we deduce that if column $1$ is covered partly by $T_2$,
then the block-tiling is exactly $\bar M^G$.  Now if row $1$ and
column $1$ are completely covered by $T_1$, then $(I_1\cup I_2)\times
J$ and $I\times (J_1\cup J_2)$ are completely covered by $T_2-$tiles.
As a result the block-tiling only contains in $(I_3\cup I_4)\times
(J_3\cup J_4)$ either $T_1-$tiles or $T_2-$tiles, that correspond with
the $\bar M^Y$ tiling and another we call the bad tiling,
respectively.

We will show that no bad tiling appears in $\bar M$.  Let $N_R$ be the
number of blocks in $\bar M$ that are $\bar M^R$.  Similarly, let
$N_B$ be number of bad block-tilings in $\bar M$.  Note that the row
projection of a bad tiling equal the row projections of $\bar M^G$ and
that the column projections equal the projections of $\bar M^R$.
Therefore by the projections we have the equalities
\begin{align*}
  N_R  &= \sum_i r^R_i
&
  N_R + N_B &= \sum_j s^R_j.
\end{align*}
Since by assumption $\sum_i r^R_i = \sum_j s^R_j$, we have $N_B=0$.
This shows the second requirement of our construction, and by
Lemma~\ref{lem:tiling-structure} completes the proof.
\end{myproof}

\subsection{An NP-hardness proof for three rectangular tiles}
\label{sec:3rect}

\begin{theorem}
  The tiling tomography problem is NP-hard for any 3 rectangular tiles.
\end{theorem}
\begin{myproof}[sketch]
  Let $p_1\times q_1$, $p_2\times q_2$ and $p_3 \times q_3$ the
  respective dimensions of 3 tiles $T_1,T_2,T_3$.  

	The idea of the construction is that we apply the general proof scheme from section~\ref{sec:scheme}
	with 3 tilings  $\bar M^R, \bar M^{G}, \bar M^{Y}$, such that $\bar M^{R}$ contains tile $T_{1}$ in position $(0,0)$, $\bar M^{G}$ contains $T_{2}$ and $\bar M^{Y}$ contains $T_{3}$ in position $(0,0)$. Moreover each of the 3 tiling minimizes lexicographically 
	 $n_{1},n_{2},n_{3}$, where $n_{c}$ is the number of tiles $T_{c}$ in the tiling.
	 
	 Formally,
  let $i_1$ be the smallest number $i>0$ with $i \bmod p_{1}=0$ and either $i \bmod p_{2}=0$ or $i \bmod p_{3}=0$.   Let $i_2$ be the
  smallest number $i\geq i_{1}$ with $(i-i_{1})\bmod p_{2}=0$ and $i \bmod p_{3}=0$.
  Let $i_{3}$ be the smallest number $i>0$ with $i \bmod p_{2} = 0$ and $i \bmod p_{3}=0$.
  We define numbers $j_1,j_2,j_3$  in exactly the same manner
  with $q_1,q_2,q_3$ playing the same role as $p_1,p_2,p_3$.

  We apply Lemma~\ref{lem:tiling-structure} for $k=\max\{i_{2},i_3\}$ and
  $\ell=\max\{j_{2},j_3\}$.  The 3 tilings of the $\ell\times k$ grid are depicted in
  figure~\ref{fig:3rectangles}, and defined formally as follows. In this section we assume for convenience that the rows and column indices relative to a block start at $0$ instead of $1$.

  In $\bar M^R$, the $[1,i_1]\times [1,j_1]$ subsquare is completely
  tiled with $T_1$.  
  Then the region $[1,i_{2}] \times [1,j_{2}] - [1,i_{1}]\times [1,j_{1}]$ 
  is completely tiled with $T_{2}$ and the
  remaining part with $T_{3}$.
  Note that by the choice of $k$, no column is intersects a tile $T_{1},T_{2}$ 
 \emph{and} $T_{3}$. For example column $0$ intersects $T_{1}$ tiles from row $0$ to $i_{1}-1$, and then either tiles $T_{3}$ from row $i_{1}$ to $k$ (if $i_{1} \bmod p_{3}=0$) or tiles $T_{3}$ from row $i_{1}$ to $k$
 (if $i_{1} \bmod p_{3}\neq 0$). The same holds for rows.

  In $\bar M^G$, the $[1,i_3]\times [1,j_3]$ subsquare is completely
  tiled with $T_2$, and the remaining part with $T_3$.
  In $\bar M^Y$, the whole $[1,k]\times [1,\ell]$ block is tiled with
  $T_3$.

  \begin{figure}
    \centerline{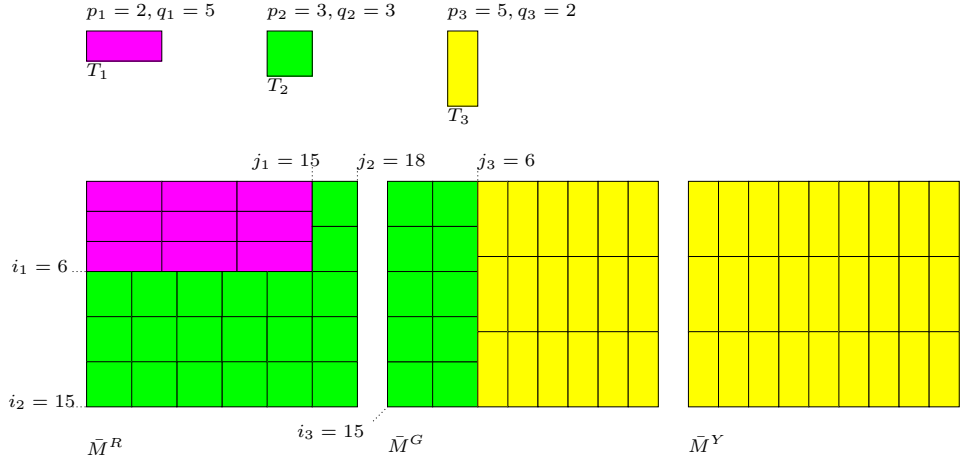}
    \caption{The 3 valid block tilings.}
    \label{fig:3rectangles}
  \end{figure}

  Clearly the projections of these tiles satisfy the first requirement
  for the general proof structure.  Now there are different
  interesting observations to make.  In the tilings above, every tile
  $T_c+(i,j)$ satisfies $i \bmod p_c = 1$ and $j \bmod q_c = 1$.
  This means that the projections of tile $T_c$ are zero for any row
  $i \neq 1 \pmod {p_c}$ or column $j \neq 1 \pmod {q_c}$, and in any
  solution $M$ to the tiling instance resulting from the reduction,
  the property must hold for all blocks as well.  This
  observation is crucial for the proof.

  In particular it implies the following fact (*). Fix some solution $M$ to the tiling instance resulting from the reduction. Consider a block in $M$. If there is some \emph{vertical separation} between two types of tiles, in the sense that cell $(i-1,j)$ is covered by some tile $T_{a}$ and $(i,j)$ by some tile $T_{b}$ with $a\neq b$, then we must have $i \bmod p_{a}=1$ and $i\bmod p_{b} =1$.

We use this observation to show that the construction satisfies also the second requirement.
  Fix some solution $M$ to the tiling instance resulting from the reduction. 
  We distinguish 3 types of blocks: (1) Blocks that contain a tile $T_{1}$, (2) blocks that do not contain any tile $T_{1}$, but contain a tile $T_{2}$, and (3) blocks that are completely tiled with $T_{3}$.
  So blocks of the third type are exactly $\bar M^{Y}$, and we have to show that blocks of the first type are exactly $\bar M^{R}$ and blocks of the second type are $\bar M^{G}$. 
  
   Let $n_{R}$ be the number of blocks of the first type, and consider one of them.
 Then by the projections all tiles $T_{1}$ must be contained in the region $[1,i_1]\times[1,j_1]$, and by the observation (*) above, the whole region must be tiled with $T_{1}$.
  Now if $i_{2}>i_{1}$, then the region $[1,i_{2}] \times [1,j_{2}] - [1,i_{1}]\times [1,j_{1}]$ cannot contain tiles $T_{1}$ nor $T_{3}$ and must be completely tiled with $T_{2}$.  Later we will show that the remaining part of the block is tiled with $T_{3}$.
  
  Note that $n_{R}$ is also the total number of red projections in the original 3-color tomography instance, so if $i_{3} \leq i_{2}$, then all $T_{2}$-tiles that have to be placed in a block-row $i
  \geq \max\{i_{2},i_{3}\}$, are placed in a type 1 block. The same observation can be made for columns. Therefore all tiles $T_{2}$ in a type 2 block, must be placed at positions of the form $(i,j) \in [1,i_{3}]\times[1,j_{3}]$. By the observation (*), the whole region $[1,i_{3}]\times[1,j_{3}]$ is completely tiled with $T_{2}$. By the observation above, the remaining part can only be tiled with $T_{3}$, which shows that the type 2 blocks are exactly $\bar M^{G}$.
  
  Let $n_{G}$ be the number of type 2 blocks in $M$.  It is also the total number of green projections in the original 3-color tomography instance. Therefore by the projections, every tile $T_{2}+(i,j)$ with $i>\max\{i_{2},i_{3}\}$ or $j>\max\{j_{2},j_{3}\}$ must be contained in a type 2 block. This shows that the remaining part $[1,k]\times [1,\ell] - [1,i_{2}]\times[1,j_{2}]$ of a type 1 block, contains only $T_{3}$-tiles. This shows that the type 1 blocks are exactly $\bar M^{R}$.
  
  Therefore the construction satisfies the second requirement for
  Lemma~\ref{lem:tiling-structure}, and we are done.
\end{myproof}

\section{Acknowledgement}

We thank Christophe Picouleau and Dominique de Werra for correcting an earlier version of this manuscript.

\bibliographystyle{plain} \bibliography{disctomo}

\begin{thebibliography}{1}

\bibitem{Alpers.Rodek.ea:Advances-in-Discrete}
A.~Alpers, L.~Rodek, H.F. Poulsen, E.~Knudsen, and G.T. Herman.
\newblock {\em Advances in Discrete Tomography and Its Applications}, chapter
  Discrete Tomography for Generating Grain Maps of Polycrystals, pages
  271--301.
\newblock Birkh\"auser Boston, 2007.

\bibitem{ChrobakDurr01}
M.~Chrobak and C.~D\"urr.
\newblock Reconstructing polyatomic structures from discrete {X}-rays:
  {NP}-completeness proof for three atoms.
\newblock {\em Theoretical Computer Science}, 259:81--98, 2001.

\bibitem{CCDW01}
Marek Chrobak, Peter Couperus, Christoph D{\"u}rr, and Gerhard Woeginger.
\newblock On tiling under tomographic constraints.
\newblock {\em Theoret. Comput. Sci.}, 290(3):2125--2136, 2003.

\bibitem{durr.goles.ea:03:tiling*}
Christoph D\"urr, Eric Goles, Ivan Rapaport, and Eric R{\'e}mila.
\newblock Tiling with bars under tomographic constraints.
\newblock {\em Theoret. Comput. Sci.}, 290(3):1317--1329, 2003.

\bibitem{GaGrPr99}
R.~Gardner, P.~Gritzmann, and D.~Prangenberg.
\newblock On the computational complexity of reconstructing lattice sets from
  their {X}-rays.
\newblock {\em Discr. Math.}, 202:45--71, 1999.

\bibitem{GarJoh79}
M.R. Garey and D.S. Johnson.
\newblock {\em Computers and Intractability: A Guide to the Theory of
  {NP}-Completeness}.
\newblock W.H.Freeman and Co., 1979.

\bibitem{Kuba.Herman:Discrete-tomography}
Attila Kuba and Gabor~T. Herman.
\newblock {\em Discrete tomography: Foundations, Algorithms and Applications},
  chapter Discrete tomography: A Historical Overview.
\newblock Birkh\"auser, 1999.

\bibitem{ryser60:_matric_zeros_ones}
H.J. Ryser.
\newblock Matrices of zeros and ones.
\newblock {\em Bull. Am. Math. Soc.}, 66:442--464, 1960.

\bibitem{thiant06:_const_et_recon_de_pavag_de_domin}
Nicolas Thiant.
\newblock {\em Constructions et reconstructions de pavages de dominos}.
\newblock PhD thesis, Universit\'e Paris 6, 2006.

\end{thebibliography}

\end{document}